\documentclass[%
reprint,
superscriptaddress,
showpacs,preprintnumbers,
 amsmath,amssymb,
 aps,
]{revtex4-1}
\usepackage{color}
\usepackage{lineno}
\usepackage{graphicx}
\usepackage{dcolumn}
\usepackage{bm}

\begin{document}

\title{Pressure induced Hydrogen-Hydrogen interaction in metallic FeH revealed by NMR}

\author{Thomas Meier}
   \email{thomas.meier@uni-bayreuth.de}
   \affiliation{Bavarian Geoinstitute, University of Bayreuth, D-95447 Bayreuth, Germany}

\author{Florian Trybel}
  \affiliation{Bavarian Geoinstitute, University of Bayreuth, D-95447 Bayreuth, Germany}    
 
\author{Saiana Khandarkhaeva}
  \affiliation{Bavarian Geoinstitute, University of Bayreuth, D-95447 Bayreuth, Germany}
   
\author{Gerd Steinle-Neumann}
  \affiliation{Bavarian Geoinstitute, University of Bayreuth, D-95447 Bayreuth, Germany}   
 
\author{Stella Chariton}
  \affiliation{Bavarian Geoinstitute, University of Bayreuth, D-95447 Bayreuth, Germany}       

\author{Timofey Fedotenko}
  \affiliation{Material Physics and Technology at Extreme Conditions, Laboratory of Crystallography, University of Bayreuth, D-95447 Bayreuth, Germany}

\author{Sylvain Petitgirard}
  \affiliation{Institute of Geochemistry and Petrology, Department of Earth Sciences, Eidgen\"ossische Technische Hochschule Z\"urich, S-8092, Switzerland}       
  
 \author{Michael Hanfland}
 \affiliation{European Synchrotron Radiation Facility (ESRF), F-38043 Grenoble Cedex, France}    
  
 \author{Konstantin Glazyrin}
 \affiliation{Deutsches Elektronen-Synchrotron (DESY), D-22603, Hamburg, Germany}    
  
\author{Natalia Dubrovinskaia}
  \affiliation{Material Physics and Technology at Extreme Conditions, Laboratory of Crystallography, University of Bayreuth, D-95447 Bayreuth, Germany}                         

\author{Leonid Dubrovinksy}
  \affiliation{Bavarian Geoinstitute, University of Bayreuth, D-95447 Bayreuth, Germany}  

\date{\today}

\begin{abstract}
Knowledge of the behavior of hydrogen in metal hydrides is the key for understanding their electronic properties. So far, no experimental methods exist to access these properties at multi-megabar pressures, at which high-$T_c$ superconductivity emerges. Here, we present an $^1$H-NMR study of cubic FeH up to 202 GPa. We observe a distinct deviation from the ideal metallic behavior between 64 and 110 GPa that suggests pressure-induced H-H interactions. Accompanying ab-initio calculations support this result, as they reveal the formation of an intercalating sublattice of electron density, which enhances the hydrogen contribution to the electronic density of states at the Fermi level. This study shows that pressure induced H-H interactions can occur in metal hydrides at much lower compression and larger H-H distances than previously thought and stimulates an alternative pathway in the search for novel high-temperature superconductors. 
\end{abstract}

\maketitle
Hydrides and hydrogen-rich compounds attract considerable attention as the search for effective and "green energy" materials intensifies \cite{Mohtadi2016, Zhang2017}. Recent theoretical \cite{Gorkov2016}, computational \cite{Zurek2009, Jarlborg2016, Bianconi2015} and experimental \cite{Drozdov2015a, Somayazulu2018} results indicate that hydrides may hold the key to a deep understanding of high-temperature superconductivity and the synthesis of compounds exhibiting high critical temperatures ($T_c$). Sulphur hydride, $H_3S$, with a $T_c$ of 200 K at pressures of about 1.5 Mbar \cite{Drozdov2015} is a prominent example of such metal hydride (MH) systems.\\ A systematic  analysis of computational results \cite{Semenok2018, Semenok2018a} suggests that two properties are particularly important for achieving high-Tc superconductivity in metallic MH: (1) significant contribution of hydrogen to the electronic density of states at the Fermi energy (EF), and (2) strong effects of hydrogen vibrations on the electronic structure of the material (i.e. electron-phonon coupling). Unfortunately, the lack of experimental methods in high pressure research able to access the electronic states of hydrogen in MHs at extreme pressures has prohibited a direct confirmation of this hypothesis.\\
Recent developments in our group led to the implementation of Nuclear Magnetic Resonance (NMR) spectroscopy in diamond anvil cells (DACs) at pressures approaching the megabar regime using magnetic flux tailoring Lenz lenses \cite{Meier2017, Meier2018, Meier2018a}. NMR spectroscopy is widely recognized \cite{Carter1977, Baek2009, Fujiwara2002a, Meier2018b} for its sensitivity to fine electronic effects. In particular, Knight shift measurements provide a well-established technique to investigate the density of states of conduction electrons at the Fermi energy, $N(E_F)$, enabling detection of deviations from free electron gas behavior or even electronic topological transitions of the Fermi surface and providing direct evidence of the Meissner-Ochsenfeld effect in the metal's superconducting state.\\
In density functional theory (DFT) based calculations Peng et al. \cite{Peng2017} demonstrated that in clathrate-like metal hydrogen systems, such as LaH$_10$ or YH$_10$, nearest H-H distances should be $\leq$ 1.5 \AA for the development of significant hydrogen-hydrogen interactions. In non-magnetic fcc FeH, stable above 25 GPa \cite{Narygina2011}, average proton-proton distances are in the range of 2.6 - 2.3 \AA~ at 200 GPa \cite{Pepin2014a}, suggesting that pressure induced H-H correlations are unlikely. As the synthesis of fcc FeH in a DAC is well established and the crystal structure well defined, hydrogenation and composition of FeH samples can be closely controlled. Therefore, fcc iron monohydride can be regarded as an ideal archetypal test or reference system for the investigation of the electronic structure of hydrogen in metal hydrides.\\
\begin{figure*}[htb]
  \includegraphics[width=1.65\columnwidth]{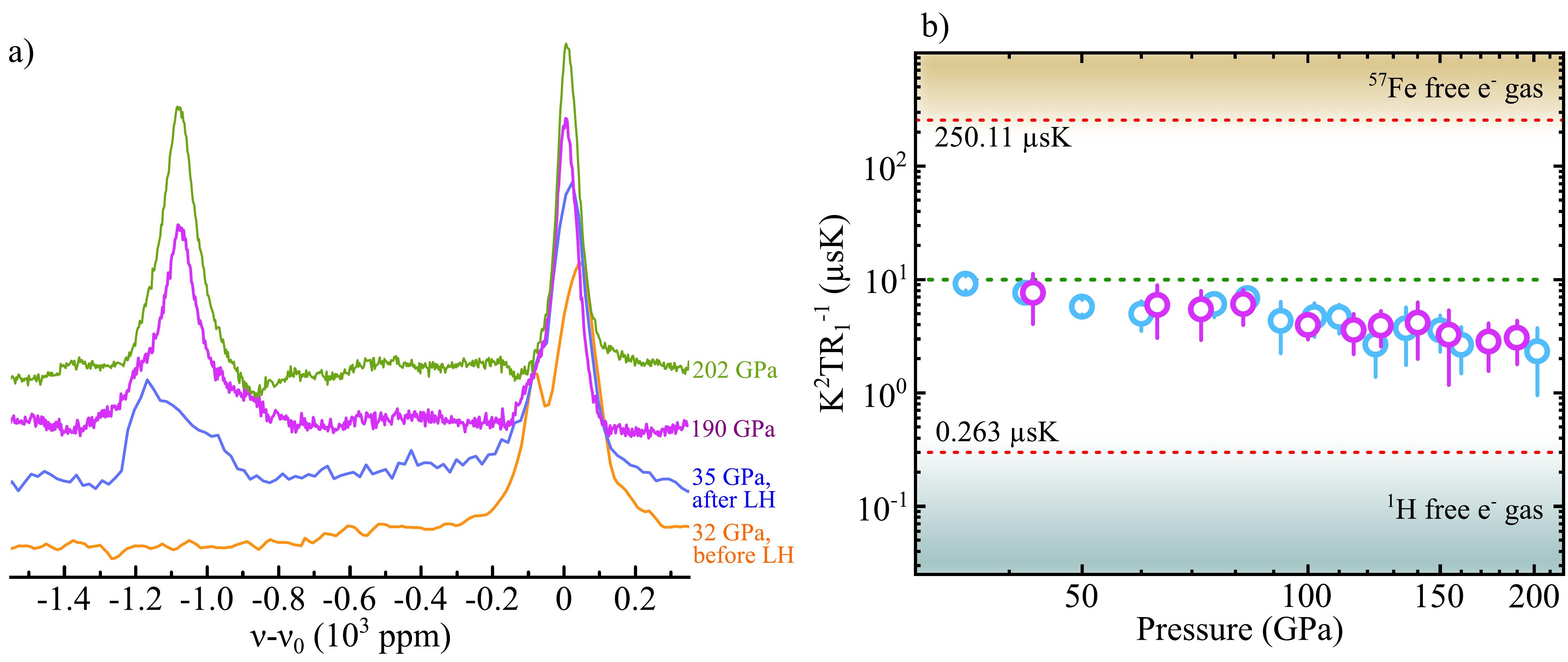}%
 \caption{Summary of experimental data. a) Proton NMR spectra. Shown are spectra before laser heating at 32 GPa (orange), directly after laser heating (blue), and after compression to 190 and 202 GPa (purple and green). b) Pressure dependence of the Korringa ratio. Shaded areas qualitatively depict the regions expected for pure iron (top) and metallic hydrogen (bottom). Light blue and purple open dots refer to experiments in cell \# 1 and cell \# 2, respectively.
 \label{FIG1}}
 \end{figure*} 
For an NMR study of FeH under pressure, an FeH sample was synthesized in a DAC at about 32 GPa through a direct reaction between iron and paraffin upon laser heating of their mixture at 1200 K. The DAC preparation, the synthesis, and experimental procedure are described in detail in the Supplementary Materials. Fig. 1a shows typical proton spectra prior to and after the sample synthesis (at about 32 GPa and 35 GPa, respectively) and after compression to 190 and 202 GPa. Before laser heating only a single NMR signal is observed, originating from the paraffin proton reservoir which has been characterized in earlier experiments \cite{Meier2017}. After the sample synthesis, an additional $^1$H-NMR signal appears at about -1200 ppm relative to the paraffin proton reservoir (Fig. 1a; Fig. S1). If this tremendous shift is due to an onset of hydrogen Knight shift ($K_H$) \cite{Knight1949} in the metallic state, nuclear relaxation, as described by Heitler and Teller \cite{Heitler1936}, must occur as well. We find, indeed, that spin-lattice relaxation rates ($R_1$) increase by more than one order of magnitude from $\approx$ 1 Hz in the proton reservoir to $\approx$ 40 Hz in the spin system associated with the additional signal (Fig. 1a; Fig. S2). Combining shift and relaxation rates, the Korringa ratio \cite{Korringa1950}, $K_H^2TR_1^{-1}$ (where $K_H$, $R_1$ and $T$ denote the hydrogen Knight shift, the spin-lattice relaxation rate and the sample temperature, respectively; see Supplementary Materials for more information), should yield values intermediate to those for pure $^{57}$Fe metal ($\approx$ 250 $\mu$sK) and Fermi-gas like value of the proton site ($\approx$ 0.3 $\mu$sK). The value of the Korringa ratio recorded directly after sample synthesis was found to be $\approx$ 10 $\mu$sK, half-way in magnitude between these two values (Fig. 1b). Therefore, the appearance of the additional signal can be attributed to the formation of metallic FeH in the DAC. Similar proton NMR spectra of other metal hydride systems, synthesized at ambient conditions, confirm this conclusion \cite{Goring1981, Goren1986}. Figure 1b also shows the pressure evolution of the Korringa ratio which should be volume independent in metallic systems \cite{Meier2015}. However, a distinct pressure dependence is observed between 30 and 202 GPa. To investigate this phenomenon, we focused on the analysis of the volume dependence of the hydrogen Knight shift $K_H$.\\
As $K_H$ scales with the density of states of d-electrons at the Fermi energy, $N_d(E_F)$, in transition metal hydrides (see Supplementary Materials for details), a volume dependence with a scaling exponent of $\alpha = 2/3$ for the free electron gas can be expected. Figure 2a shows a double logarithmic power plot of the relative volumes, using the equation of state (EOS) from Narygina et al. \cite{Narygina2011}, against relative changes in the Knight shift. Three distinct regions can be identified: 1) up to 64 GPa, the exponent closely follows a Fermi-gas electronic behavior; 2) a deviation from metallic scaling exponents occurs between 64 GPa and 110 GPa and 3) the re-occurrence of free electron gas behavior from 110 GPa to highest pressures. After compression and data collection up to 202 GPa, one of the DAC experiments was decompressed to $\approx$ 100 GPa and subsequently to 33 GPa (red arrows in Fig. 2a). NMR spectra collected on decompression agree with those obtained during compression, showing that the observed effect is fully reversible. As X-ray diffraction data do not show any sign of a structural phase transition (see Fig. S3a and  \cite{Pepin2014a, Thompson2018}), the origin of this observation should be linked to changes in the electronic environment of the protons in FeH. \\
\begin{figure*}[htb]
  \includegraphics[width=1.5\columnwidth]{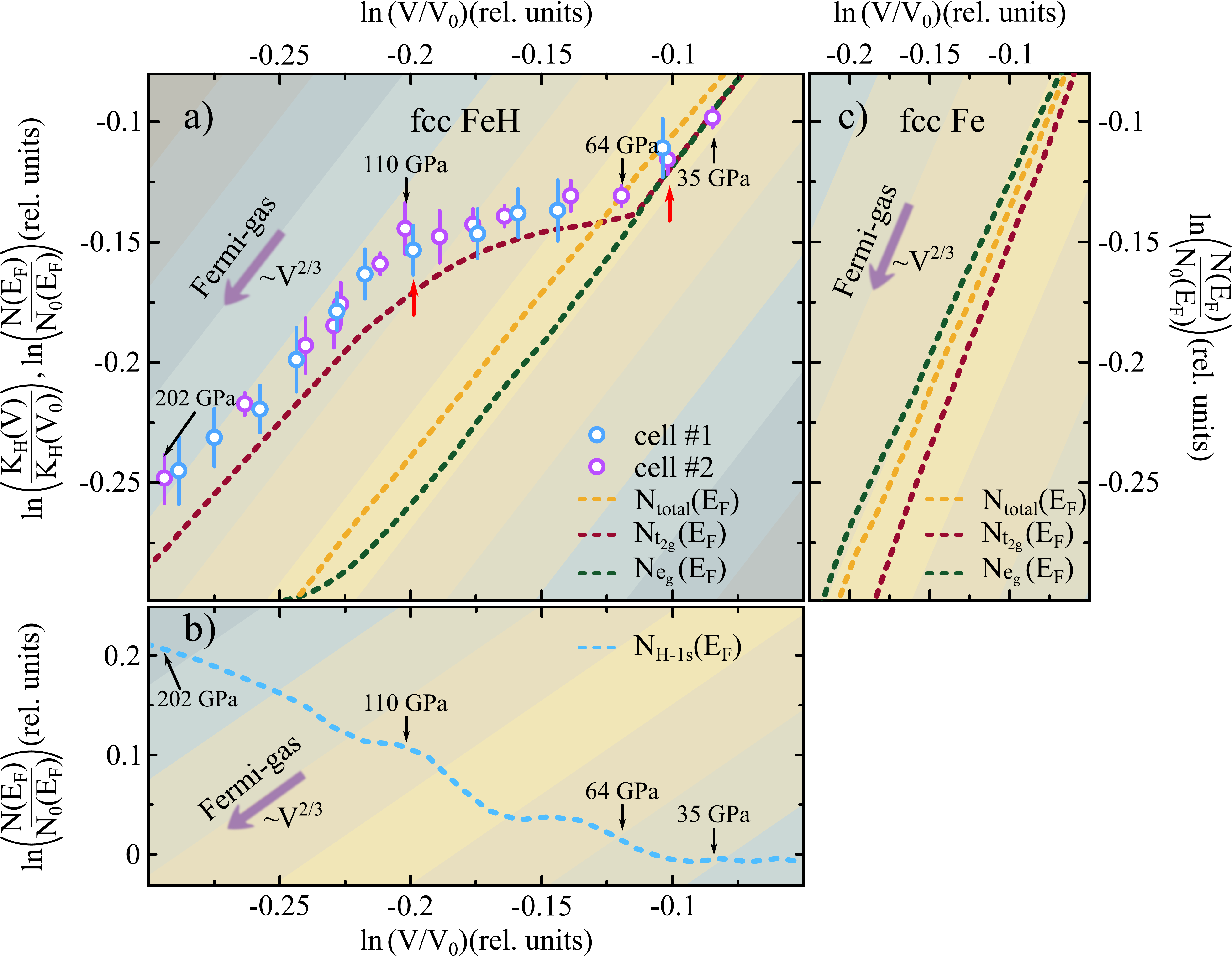}%
 \caption{Comparison of experimental data and ab-initio electronic DOS calculations. a)  Double logarithmic power plot of relative changes in $K_H$ and $N(E_F)$ as a function of the relative volume. Experimental data points, blue and purple, have been normalized to $V_0$ using the EOS from Narygina et al. and extrapolated values of $K_H(V_0)$. Note that using complementary EOS data from Thompson et al.\cite{Thompson2018} or Pepin et al. \cite{Pepin2014a} did not affect our conclusions. The dashed lines (splines through computed values) show the dependence of $N_{total}(E_F)$, $N_{t2g}(E_F)$ and $N_{eg}(E_F)$. The diagonal color strips are guides to the eyes depicting a  $\propto V^{2/3}$ scaling for free electron Fermi-gas like behavior. Black arrows denote respective pressures points; red arrows show data recorded under decompression. b) Evolution of $N_{H-1s}(E_F)$. Indicated pressure markers are related to the experimental data points from a). c) Pressure dependence of $N_{total}(E_F)$, $N_{t2g}(E_F)$ and $N_{eg}(E_F)$ for fcc Fe in a similar pressure range.
 \label{FIG2}}
 \end{figure*} 
To explore this hypothesis, we applied Kohn-Sham DFT using Quantum Espresso \cite{Giannozzi2009} (see Supplementary Materials for computational details). Both electronic band structure and density of states (Fig. S4) do not indicate any discontinuities such as van-Hove singularities \cite{Volovik2017} close to $E_F$. Therefore topological transitions of the Fermi surface \cite{Lifshitz1960}, which would influence both spin shift and relaxation rates, can be excluded. Values of the total density of states at $E_F$, $N_{total}(E_F)$, as well as its projection on the $3d~t2g$ and $eg$ states, $N_{t2g}(E_F)$ and $N_{eg}(E_F)$, respectively, from the calculations are included in Fig. 2a (dashed lines). The volume dependence of $N_{t2g}(E_F)$ closely follows the NMR spin shifts to 64 GPa. Above this pressure, experimental data points and $N_{t2g}(E_F)$ diverge beyond experimental errors. \\
The hydrogen Knight shift is the sum of interactions due to core polarization of uncompensated d-orbitals and contact interaction of s-electrons with the nuclei (see equation (3) in the Supplementary Materials). At ambient conditions, the latter interaction can be regarded negligible due to the density of states at $E_F$ being almost zero for the hydrogen 1s-electron, $N_{H-1s}(E_F)$, (Fig. 2b). The contribution of $N_{H-1s}(E_F)$ to the total DOS begins to increase around 64 GPa and continuously rises up to 200 GPa. This effect qualitatively explains the observed offset between experimental data and calculated DOS values.  \\
Figure 2c shows the respective volume dependence of the density of states at $E_F$ of the 3d electrons for fcc-Fe over the same compressional range as the FeH data. The pressure dependent effects observed on the DOS of the 3d-t2g orbitals in fcc FeH, seen in Fig. 2a, are not present in fcc Fe. That allows us to conclude that they are caused by the presence of hydrogen atoms. Both the scaling anomaly in $N_{t2g}(E_F)$ and the continuous increase in $N_{H-1s}(E_F)$ with compression suggest a shift of conduction electron density between iron and hydrogen atoms in FeH. 
\begin{figure*}[htb]
  \includegraphics[width=1.5\columnwidth]{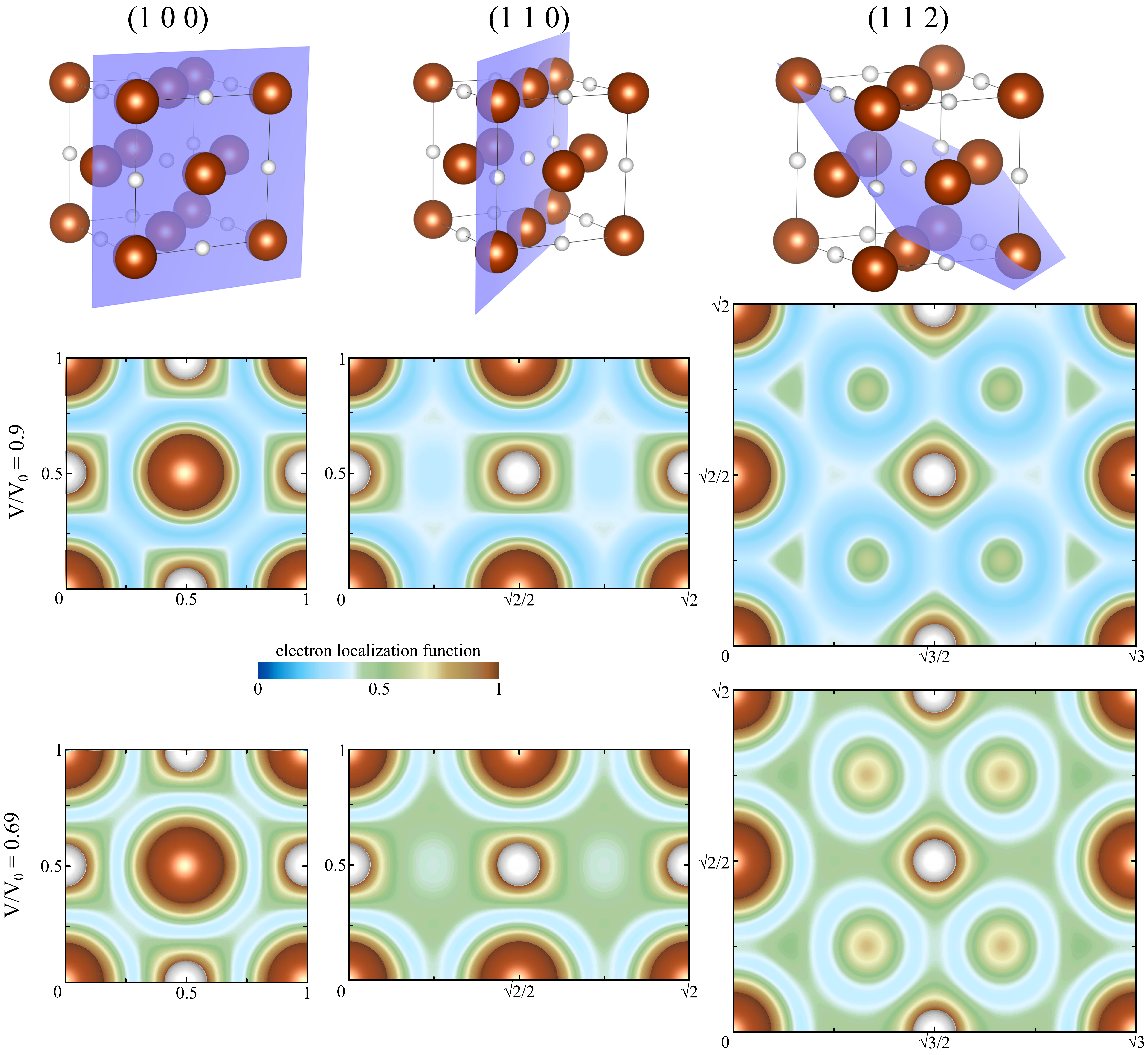}%
 \caption{Electron localization function maps in three major crystal planes at a compression of $V/V_0$=0.9 and 0.69, corresponding to 30 and 200 GPa. Brown spheres represent iron atoms, white spheres hydrogen. Green areas (ELF $\approx$ 0.5) correspond to Fermi-gas like regions of free electrons
 \label{FIG2}}
 \end{figure*}
In order to elucidate this possibility, electron localization function (ELF) maps have been computed. Fig. 3 shows the ELF maps in three different crystal planes at compressions of $V/V_0$=0.9 and  $V/V_0$=0.69 corresponding to about 30 and 200 GPa. At  $V/V_0$=0.9, electron clouds associated with Fe and H atoms are localized, evidencing the ionic character of FeH. Regions exhibiting Fermi-gas electronic behavior, i.e. ELF values of 0.5, start to develop and spread at  $V/V_0~<$ 0.88, which are particularly pronounced in the (110) and (112) planes. These regions of increased electron density connect at $V/V_0~<$ 0.78, corresponding to pressures of 80-90 GPa, forming an interconnecting network with respect to the hydrogen nuclei. At $V/V_0~<$ 0.73, the nearest neighbor hydrogen-hydrogen bonds are bridged as well. An animation of the results (Movie S1) shows the correlation between the different ELF maps and the data shown in Fig. 2. Strikingly, the deviation from free electron gas behavior observed for $V/V_0~=$0.88 coincides with the increase of electron density in the interstitial regions of FeH. The reemergence of Fermi-gas behavior above 110 GPa corresponds to the connection of these areas of enhanced electron density. The pronounced increase in $N_{H-1s}(E_F)$ also correlates qualitatively with both the formation and connection of these areas. 
Combining experimental observations and computational results, we propose the following qualitative explanation of the high-pressure behavior of FeH: With increasing pressure, in FeH electron density from partially uncompensated 3d-t2g Fe orbitals is pushed towards the hydrogen electron clouds. The successive addition of charges leads to the formation of an intercalating free-electron sub-lattice of connected hydrogen atoms, exhibiting a continuous enhancement of the electronic density of states from the hydrogen at $E_F$. Our data suggest that this effect can be observed by NMR when average H-H distances become smaller than $\approx$ 2.5 Å which is significantly larger than expected based on previous computational work [12].
In this study of hydrogen in iron monohydride FeH we have shown that pressure induced interactions between hydrogen atoms and the transition metal atoms as well as between H-atoms themselves can occur and influence electronic properties of metal hydrides at much lower pressures than previously anticipated. The experimental findings, supported by ab-initio calculations, suggest that other candidates for phonon-mediated high temperature super-conductivity may be found far off the proclaimed "lability belt", including the rare-earth sodalite-clathrate super-hydrides \cite{Heil2019}.        
\section{Acknowledgements}
We thank Nobuyoshi Miyajima and Katharina Marquardt for help with the FIB milling. We also thank Caterina Melai, Serena Dominijanni and Catherine McCammon for help with XRD measurements.
\section{Funding}
N.D. and L.D. thank the German Research Foundation (Deutsche Forschungsgemeinschaft, DFG, projects DU 954-11/1, DU-393/13-1, and DU-393/9-2) and the Federal Ministry of Education and Research, Germany (BMBF, grant no. 5K16WC1) for financial support. S.P., and F.T. were financed by the German Research Foundation (DFG projects PE 2334/1-1, FOR 2440).
\section{Author Contributions}
T.M. and L.D. designed the experiment. T.M., S.K. and S.P. prepared the DACs and NMR resonators. S.K. performed focused ion beam cutting as well as structure refinement after laser heating. L.D. performed the sample synthesis in the DACs. S.C. and T.F. as well as S.D. and C.M. performed XRD measurements at DESY and ESRF, respectively. F.T. and G.S.N. performed the calculations and analyzed the results. T.M., F.T., G.S.N., N.D. and L.D. performed the data analysis and wrote the manuscript. 
\section{Competing Interests}
The authors declare that they have no competing interests.
\section{Data availability}
The data supporting the findings of this study are available from the corresponding author upon reasonable request.


\end{document}


\beginsupplement


\title{Pressure induced Hydrogen-Hydrogen interaction in metallic FeH revealed by NMR\\
Supplementary Material}

\author{Thomas Meier}
   \email{thomas.meier@uni-bayreuth.de}
   \affiliation{Bayerisches Geoinstitut, University of Bayreuth, D-95440 Bayreuth, Germany}

\author{Florian Trybel}
  \affiliation{Bayerisches Geoinstitut, University of Bayreuth, D-95440 Bayreuth, Germany}    
 
\author{Saiana Khandarkhaeva}
  \affiliation{Bayerisches Geoinstitut, University of Bayreuth, D-95440 Bayreuth, Germany}
   
\author{Gerd Steinle-Neumann}
  \affiliation{Bayerisches Geoinstitut, University of Bayreuth, D-95440 Bayreuth, Germany}   
 
\author{Stella Chariton}
  \affiliation{Bayerisches Geoinstitut, University of Bayreuth, D-95440 Bayreuth, Germany}       

\author{Timofey Fedotenko}
  \affiliation{Material Physics and Technology at Extreme Conditions, Laboratory of Crystallography, University of Bayreuth, D-95440 Bayreuth, Germany}

\author{Sylvain Petitgirard}
  \affiliation{Institute of Geochemistry and Petrology, Department of Earth Sciences, Eidgen\"ossische Technische Hochschule Z\"urich, CH-8092 Z\"urich, Switzerland}       
  
 \author{Michael Hanfland}
 \affiliation{European Synchrotron Radiation Facility (ESRF), F-38043 Grenoble Cedex, France}    
  
 \author{Konstantin Glazyrin}
 \affiliation{Deutsches Elektronen-Synchrotron (DESY), D-22603 Hamburg, Germany}    
  
\author{Natalia Dubrovinskaia}
  \affiliation{Material Physics and Technology at Extreme Conditions, Laboratory of Crystallography, University of Bayreuth, D-95440 Bayreuth, Germany}                         

\author{Leonid Dubrovinksy}
  \affiliation{Bayerisches Geoinstitut, University of Bayreuth, D-95440 Bayreuth, Germany}  

\date{\today}

\maketitle

\section*{Experimental Set-up}

For the experiments two identical diamod anvil cells (DAC) of type BX90~\cite{Kantor2012} were prepared with two beveled diamond anvils with 100 $\mu$m culets. Rhenium blanks were indented to about 15 $\mu$m thickness and 40 $\mu$m holes were laser cut in the flat face of the pre-indentation to form the sample gasket. Using physical and chemical vapor deposition techniques, the anvils have been covered by a 1 $\mu$m thick layer of copper and the metallic rhenium gaskets were sputtered with a 500 nm thick layer of Al$_2$O$_3$ for electrical insulation. Subsequently, using a focused ion beam, three Lenz lenses (LL) were cut out of the copper layers covering the diamond anvils: 1) along the anvil`s pavilion, 2) on the beveled area and 3) on the 100 $\mu$m culets. This procedure is necessary to avoid damage due to sharp edges from the 8$^{\circ}$ bevel and the culet rim when cupping under compression occurs. The nuclear magnetic resonators used in this study can be regarded as a generalization of the recently introduced double-stage Lenz lenses~\cite{Meier2018a} for single-beveled diamond anvils with culet sizes $\le$ 100 $\mu$m.
 
The cell was loaded by flooding the sample chamber with paraffin oil and then adding fine iron powder of natural composition (Sigma-Aldrich, 4N purity). Pressure in the sample chamber was estimated by the frequency shift of the first-order Raman peak of the diamond at its edge in the center of the sample cavity~\cite{Akahama2004, Akahama2006}. Figure \ref{SupplFIG3}b shows the measured Raman spectra in a pressure range from 40 GPa to 190 GPa.

The radio-frequency (RF) excitation coils were made using two identical 3 mm coils consisting of 5 turns of 200 $\mu$m teflon-insulated copper wire. The coils were fixed around the diamonds on the respective backing plates and connected to form a Helmholtz coil, ensuring optimal inductive coupling into the LL resonator structures.
Nuclear magnetic resonance (NMR) measurements were conducted at a magnetic field of 1 T using a sweepable electron spin resonance magnet with 50 mm magnet pole distance and a Techmac Redstone spectrometer for solid-state NMR applications. Throughout this study, only non-selective 90$^{\circ}$ pulses were used to acquire spectra in the frequency domain. NMR measurements prior to cell loading did not exhibit any detectable hydrogen signals within typical experimental parameters, showing the absence of spurious hydrogen signals stemming from other parts of the NMR probe. 

From RF nutation experiments, an optimal 90$^{\circ}$ pulse of 4 $\mu$s at an average pulse power of 40 W was found. This value did not change over the course of the  experiment and was found valid for both DACs used.  From the relation $B_1=\pi/(2\gamma_nt_{\pi/2})$ a $B_1$ field of 4.7 mT was deduced. Using the \textit{femm 4.2} package \cite{Meeker2018}, the high frequency $B_1$ field distribution across the 40 $\mu$m x 15 $\mu$m  sample cavity was calculated using experimental parameters similar to the nutation experiments. An average $B_1$ field of 4.8 mT was found in excellent agreement with the nutation experiments. For measurements of spin-lattice relaxation rates, a saturation recovery pulse train consisting of 16 consecutive 90$^{\circ}$ pulses was used.  

\begin{figure}[htb]
  \includegraphics[width=0.55\columnwidth]{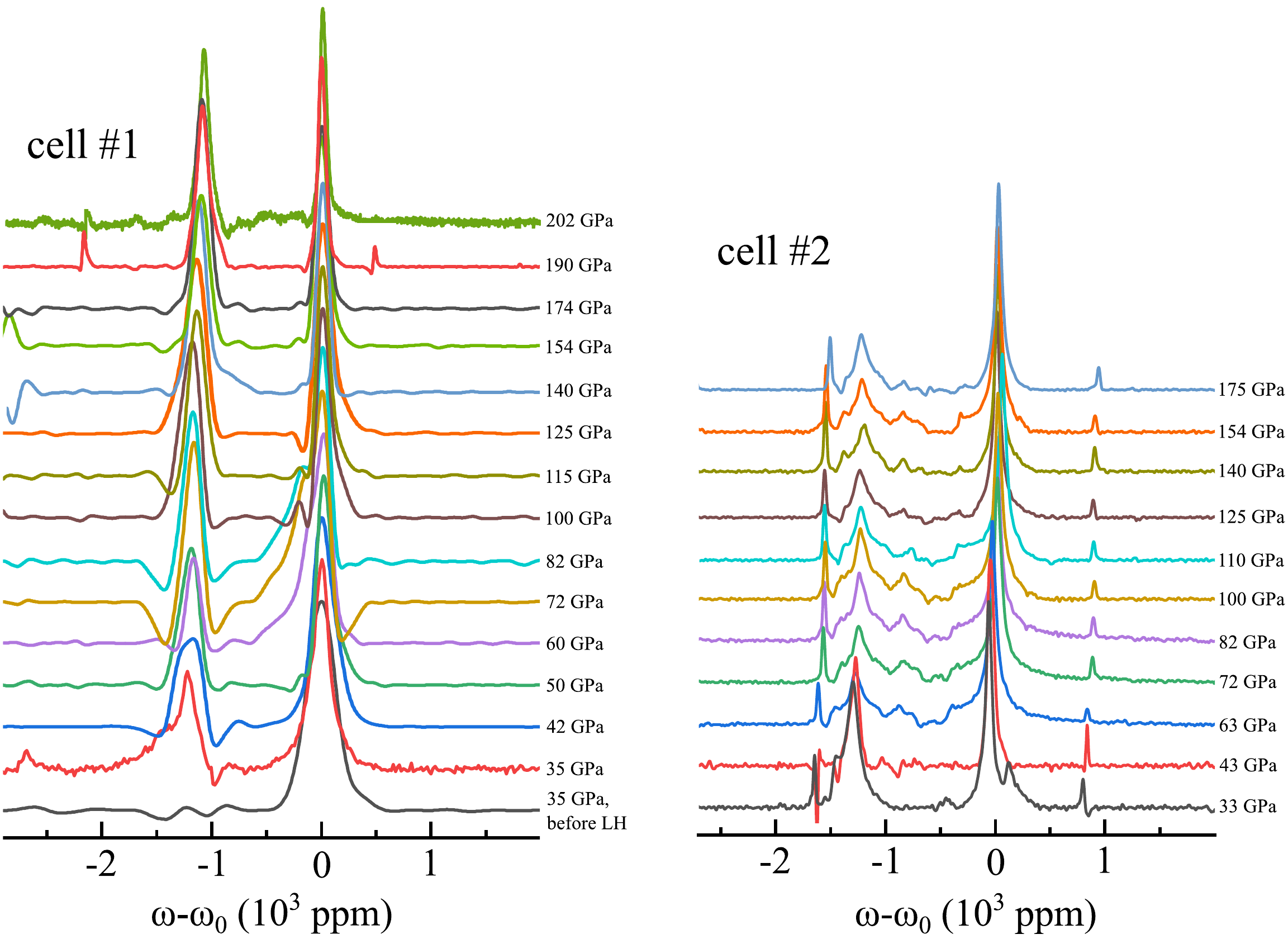}%
 \caption{$^1$H-NMR spectra of FeH and the paraffin reservoir in a pressure range $30-200$ GPa for cell \#1 (left) and cell \#2 (right).
 \label{SupplFIG1}}
 \end{figure}

\begin{figure}[htb]
  \includegraphics[width=0.50\columnwidth]{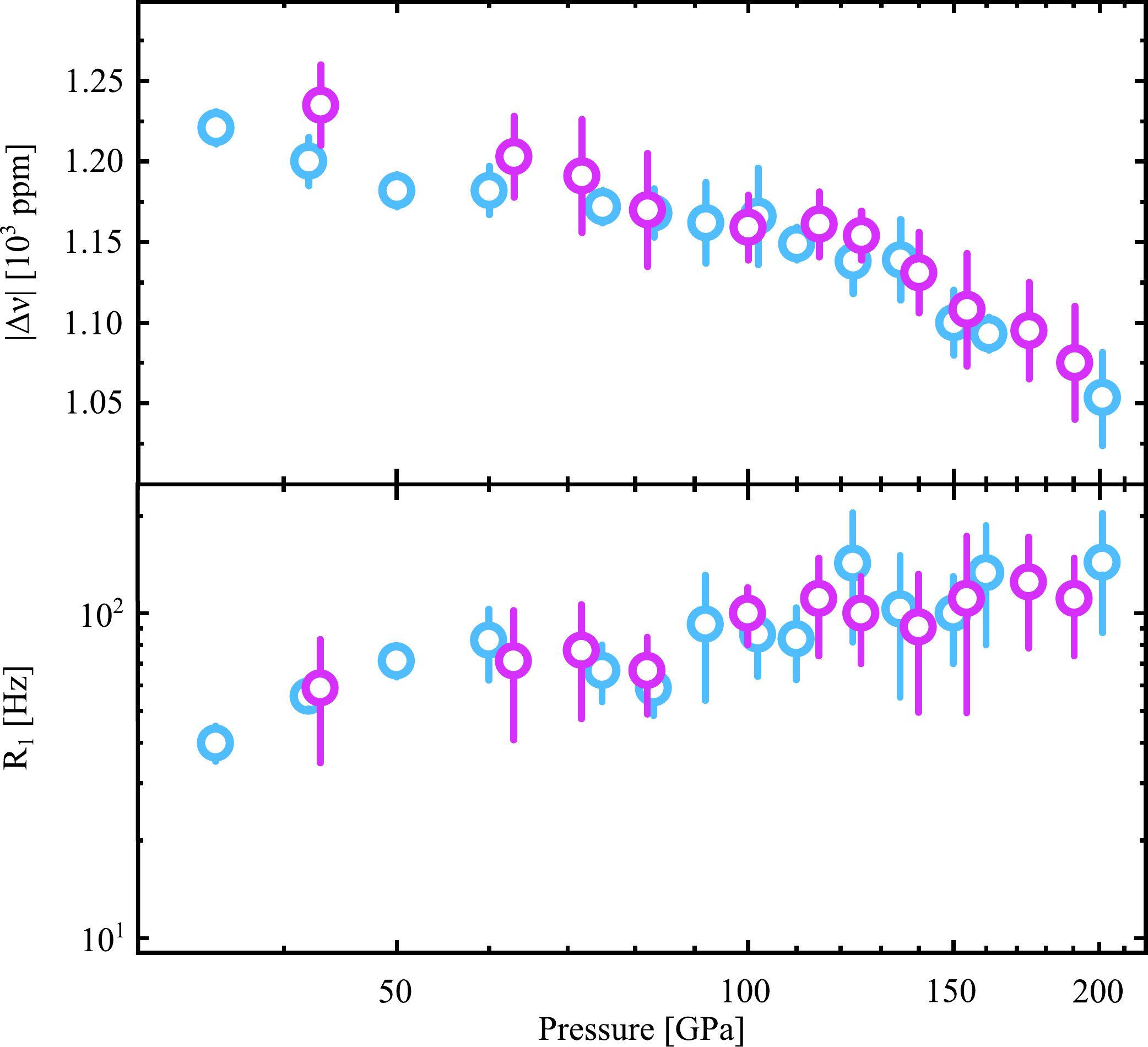}%
 \caption{ Summary of Knight shift $K_{H}$ relative to the hydrogen reservoir (top) and the spin-lattice relaxation rates R$_1$ (bottom). 
Blue and purple data points refer to data collected in cells \# 1 and \# 2 respectively 
\label{SupplFIG2}}
 \end{figure}

\begin{figure}[htb]
  \includegraphics[width=0.85\columnwidth]{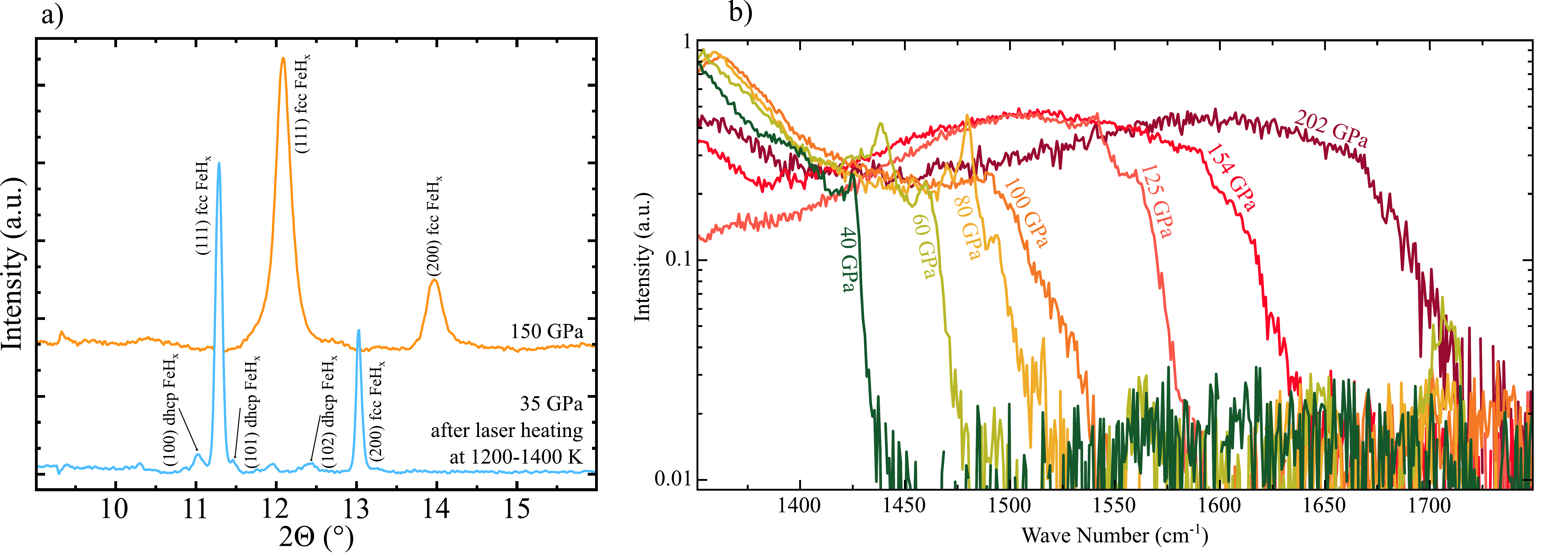}%
\caption{a) X-ray diffraction patterns after sample synthesis (laser-heating) at 35 GPa (blue) and after compression to 150 GPa (orange). b) Raman spectra of the diamond edge at the center of the 100 $\mu$m culets used for pressure calibration.
\label{SupplFIG3}}
 \end{figure}

All shifts were referenced relative to the hydrogen signal of the paraffin reservoir after laser heating. The values of the Knigh shift $K_{H}$ (Figures \ref{SupplFIG1} and \ref{SupplFIG2}) are higher by a factor of six with respect to other transition metal hydrides~\cite{Conradi2007}. The choice of the reference material can in principle lead to an intrinsic overestimate as it differs from the standard shift reference, tetramethylsilane (TMS)~\cite{Maniara1998}. However, TMS does not induce the reaction to form FeH at high pressure and temperature, to the best of our knowledge. In addition, no systematic studies on the electronic properties of laser heated paraffin have been undertaken to this date.

Two DACs were pressurized initially to about 30 GPa  (Figure \ref{SupplFIG3}b for the Raman spectra used for pressure determination) into the stability field of non-magnetic hcp Fe, and subsequently laser-heated to $1200$ K. X-ray diffraction (XRD) powder patterns taken at the extreme condition beamline (P02) at  PETRA III and the ESRF (ID-15) show reflections from FeH and no residual Fe signal (Supplementary Figure \ref{SupplFIG3}a), suggesting that iron has fully reacted with hydrogen from the paraffin reservoir. The most prominent diffraction peaks are those of fcc FeH previously identified~\cite{Narygina2011, Thompson2018}, a small amount of dhcp FeH$_x$ was also indexed. As fcc FeH$_x$ is unstable at ambient conditions, the actual chemical composition cannot be measured directly using recovered samples. Narygina et al.~\cite{Narygina2011} suggested that the hydrogenation factor $x$ can be estimated by considering the volume expansion of the fcc unit cell in response to the dissolution of hydrogen on the interstitial lattice sites of fcc Fe by $x=(V_{\mathrm{FeH_x}}-V_{\mathrm{Fe}}) / \Delta V_{\mathrm{H}}$ with V$_{\mathrm{FeH_x}}$ and $V_{\mathrm{Fe}}$ the atomic volumes of iron hydride and fcc iron, respectively. $\Delta V_{\mathrm{H}}$ is the volume expansion due to hydrogenation. As the latter parameter is not well defined by other fcc-structured $3d$ transition metals at ambient conditions, the actual hydrogenation value $x$ was estimated by comparing the recorded data with equations of state of FeH~\cite{Narygina2011} and of fcc iron~\cite{Komabayashi2010}. Using this procedure, we estimated $x=1.0(1)$ in both DACs. Further XRD measurements at 150 GPa suggested that fcc FeH remains stable with similar stochiometry throughout the whole pressure range of this study.

Prior to laser heating, $^1$H-NMR spectra were recorded (Figure \ref{SupplFIG1} and Figure \ref{SupplFIG1}): As the origin of the hydrogen signal is the C$_n$H$_{2n+2}$ reservoir, a single NMR signal positioned at the Larmor frequency for hydrogen is observed. The center of gravity of the paraffin signal was found at 45,0737 MHz, corresponding to a polarising magnetic field of $B_0=1.058$ mT within the sample cavity. Small deviations of up to 10\% in $B_0$ were found to originate from a slightly different positioning of the DACs within the region of the highest magnetic field of the electromagnet, but can be neglected since the hydrogen reservoir served as an internal reference. The recorded spectra of paraffin at 32 GPa (cell \#1) and 29 GPa (cell \#2) were found to be similar to earlier $^1$H-NMR spectra of C$_n$H$_{2n+2}$ in terms of line-widths and relaxation rates~\cite{Meier2017}. After laser heating, an additional intense signal was found at lower frequencies at about -1200 ppm relative to the hydrogen reference. Comparing signal intensities from spectra recorded after full relaxation of both independent spin systems, we observed an intensity ratio of about 4:7 of  FeH to C$_n$H$_{2n+2}$ in cell \#1 and 1:5 in cell \#2. These ratios were found to be constant at all pressures (Figure \ref{SupplFIG5}), suggesting that further reaction of hydrogen with the iron powder during the cold compression cycles did not occur.

\section*{Knight Shift in Transition Metal Hydrides}

As both stable iron hydride phases, fcc and dhcp, are thought to be metallic~\cite{Elsasser1998, Elsasser1998a}, the influence of hyperfine interaction of conduction electrons with the nuclei must be taken into account to explain the NMR signal. 
For transition metal hydrides, the magnetic susceptibility can be written as a sum of several contributions, i.e. $\chi=\chi_s+\chi_d+\chi_o+\chi_{\mathrm{dia}}$, stemming from spin interactions with $s$- or $d$-conduction electrons, the orbital paramagnetic and the diamagnetic contribution of the ion cores respectively. Therefore, the electron-nuclear hyperfine interaction, described in terms of the Knight shift $K_{\mathrm{H}}$ and the spin-lattice relaxation rate $R_1$ \cite{Knight1949, Townes1950}, are caused by various mechanisms~\cite{Goring1981}:
\begin{equation}
K_{\mathrm{H}}=K_s+K_d+K_o
=\frac{1}{A\mu_B}\Bigg(H_{\mathrm{hf}}^s \chi_s+H_{\mathrm{hf}}^d\chi_d+H_{\mathrm{hf}}^o\chi_o\Bigg)
\end{equation}
and
\begin{equation}
R_1=4\pi\mu_0^2\gamma_n^2\hbar k_BT \Bigg \lbrace [H_{\mathrm{hf}}^sN_s(E_F)]^2
+[H_{\mathrm{hf}}^dN_d(E_F)]^2q+[H_{\mathrm{hf}}^oN_d(E_F)]^2p\Bigg\rbrace, 
\end{equation}
where $A$ is Avogadro's constant, $\mu_B$ the Bohr magneton and $\gamma_n$ the gyromagnetic ratio. The hyperfine fields $H_{\mathrm{hf}}^s,~H_{\mathrm{hf}}^d$ and $H_{\mathrm{hf}}^o$ at the nucleus under investigation originate from the following mechanisms: 1) contact interaction of $s$ conduction electrons with the nucleus; 2) core polarisation of $s$-orbitals caused by uncompensated $d$ electrons and 3) the effect of orbital motion of $d$ electrons. $N_s(E_F)$ and $N_d(E_F)$ in equation (2) are the $s$- and $d$-electron density of states at the Fermi energy, respectively. The factors $p$ and $q$ are reduction factors depending on $d$-orbital degeneracy at $E_F$. 

For the free electron Fermi gas we can substitute $\chi_{(s,d)}=2\mu_0\mu_B^2N_{(s,d)}(E_F)$ in equation (1), leading to a reduced representation of the Knight shift:
\begin{align}
K_{\mathrm{H}}=2\mu_0\mu_B&\Bigg[H_{\mathrm{hf}}^sN_s(E_F)+H_{\mathrm{hf}}^dN_d(E_F)\Bigg]+\frac{1}{A\mu_B}H_{\mathrm{hf}}^o\chi_o. 
\end{align}
Equation (3) demonstrates that various mechanisms producing hyperfine fields at the hydrogen site need to be taken into account. However, as the 1$s$ electronic states of hydrogen and the 4$s$ electronic states of iron typically lie far below and above the Fermi energy, respectively, it seems reasonable to assume their contribution to (3) to be negligible to first approximation in FeH.  

The effect of orbital motion of $3d$ electrons on the hydrogen Knight shift is expected to be virtually non-exsistent as such an interaction would dominate equation (3) and lead to positive Knight shifts which is not observed in other MH systems measured at ambient conditions~\cite{Goring1981, Kazama1977a,  Kazama1977, Schreiber1965}. Also, considering the good correlation between $K_H$ and $N_{t2g}(E_F)$ (Figure 2), an influence of spin-orbit coupling on the $^1H$-NMR spectra can be ruled out. Therefore, both equations (2) and (3) are reduced to their $d$-electron core polarization contribution, with a Korringa-type~\cite{Korringa1950} relation given by:
\begin{equation}
\frac{K_H^2T}{R_1}=\frac{\hbar}{4\pi k_B} \left( \frac{\gamma_e}{\gamma_n}\right)^2q
\label{Korringa}
\end{equation}
For cubic lattices, the reduction factor $q$ can be described by $q=\frac{1}{3}f^2+\frac{1}{2}(1-f)^2$  where $f$ is the fractional character of $t_{2g}$ $d$ orbitals at the Fermi surface~\cite{Yafet1964, Narath1968}, determining the deviation from the free electron value given by $S=\frac{\hbar}{4\pi k_B} \left( \frac{\gamma_e}{\gamma_n}\right)^2$~\cite{Carter1977}. $\gamma_e$ and $\gamma_n$ are the gyromagnetic ratios of the electron and the NMR nucleus respectively. 

In general, the Korringa-ratio constitutes an indicator for metallicity of a given material~\cite{Carter1977} using the independence of equation (4) on volume.
As the Korringa-ratio only depends on physical constants, values for the free electron gas behavior for every metallic system can be readily computed. For FeH, two different scenarios can be identified: 1) pure $^{57}$Fe metal with ${K_H^2T}/{R_1}=250.11$~$\mu$sK and 2) electronic Fermi gas-like behavior of metallic hydrogen with a value of $0.263$ $\mu$sK. The Korringa-ratio for the binary system consisting of Fe and H are expected to lie in-between these extreme values.

\section*{{\it Ab-initio} calculations}

\begin{figure}[htb]
  \includegraphics[width=0.95\columnwidth]{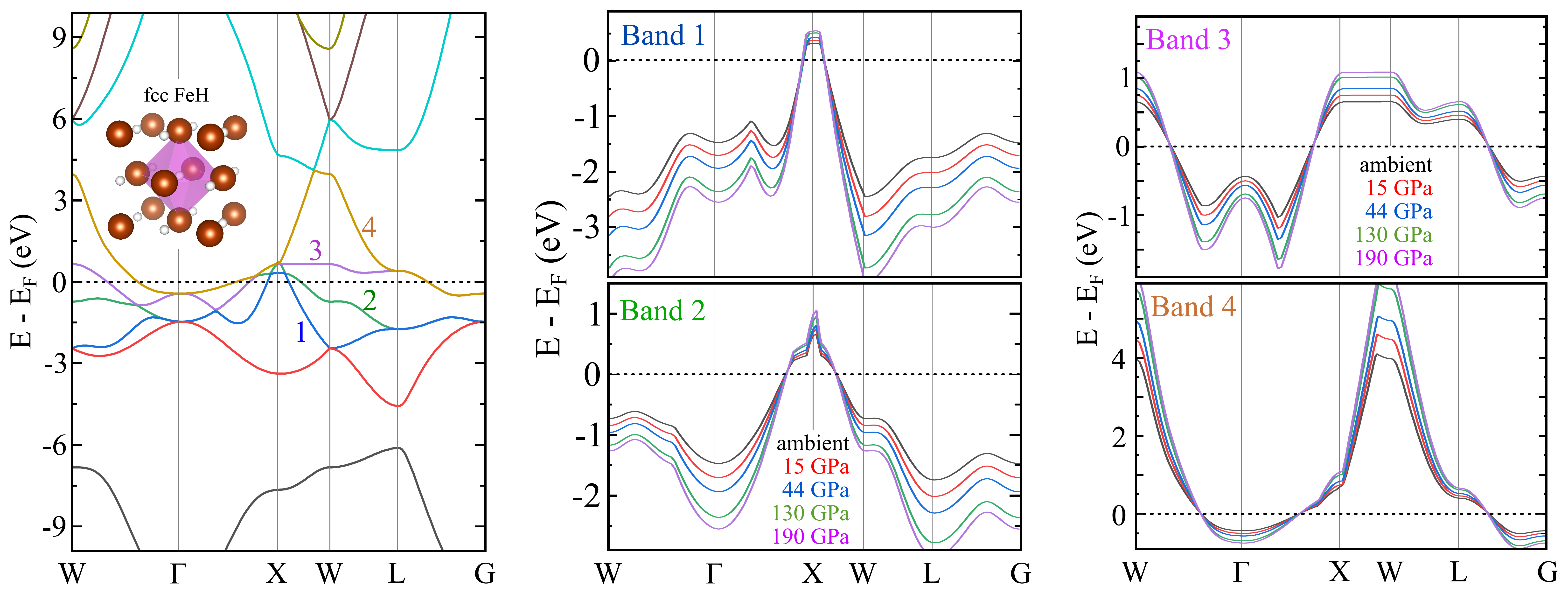}%
 \caption{a) Calculated band structure of fcc FeH. The left panel shows the band structure at ambient pressure, the two right panels the pressure evolution of four bands crossing the Fermi energy ($E_F$), identifed on the left.
 \label{SupplFIG4}}
 \end{figure}

\begin{figure}[htb]
  \includegraphics[width=0.9\columnwidth]{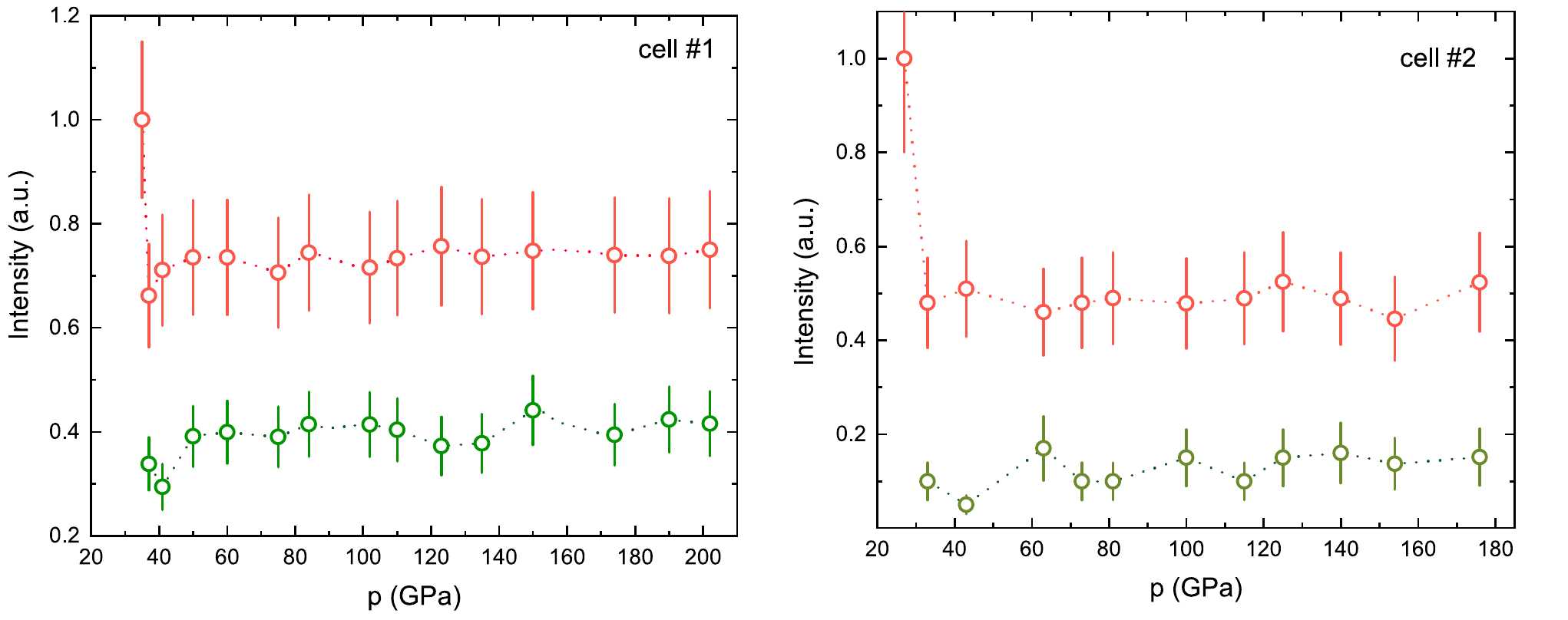}%
 \caption{Relative hydrogen NMR signal intensitites of the paraffin reservoir (red) and the signal from FeH (green) for both DAC experiments. 
 \label{SupplFIG5}}
 \end{figure}

All calculations are based on Kohn-Sham density functional theory and were performed with {\sc Quantum Espresso}~\cite{Giannozzi2009, Giannozzi2017} using the projector augmented wave (PAW) approach~\cite{Blochl1994}. We used the generalized gradient approximation by Perdew-Burke-Ernzerhof~\cite{Perdew1996} to exchange and correlation with corresponding potential files; for Fe we used a valence electron configuration that includes electronic states $3s$ and higher, applicable to extreme pressure~\cite{Stixrude2012}. Convergence tests for the electronic band structure (Figure \ref{SupplFIG4}) and density of state calculations led to reciprocal space sampling with a Monkhorst-Pack~\cite{Monkhorst1976} $k$-point grid of 32$\times$32$\times$32 and a cutoff energy for the plane wave expansion of 120 Ry. 

We explored the energetics of incorporating H on the octahedral void in the fcc Fe lattice as well as on $1/2$ of the tetrahedral voids. The former is favored energetically by $> 1$ eV per formula unit (pfu) over whole compression range explored here, i.e. $32-54$ \AA$^3$ pfu.

In order to identify regions of charge localization, we use the electron localization function~\cite{Becke1990} that quantifies the probability of finding two electrons in close proximity by the ratio of a computed charge density and its spatial derivative to the value of the homogeneous electron gas for the same density. 

\newpage

\stopsupplement
 
\clearpage